\documentclass[sigconf,prologue,table,xcdraw]{acmart}

\usepackage{multirow}
\usepackage{graphicx}
\usepackage{booktabs}
\usepackage{multirow}
\usepackage{caption}
\usepackage{makecell}
\usepackage[ruled,vlined,linesnumbered]{algorithm2e}
\usepackage{colortbl} 

\newcommand{\modelname}{EDPC}
\AtBeginDocument{%
  }

\setcopyright{rightsretained}
\copyrightyear{2025}
\acmYear{2025}
\acmConference[MM '25]{Proceedings of the 33rd ACM International Conference
on Multimedia}{October 27--31, 2025}{Dublin, Ireland}
\acmBooktitle{Proceedings of the 33rd ACM International Conference on
Multimedia (MM '25), October 27--31, 2025, Dublin,
Ireland}
\acmDOI{10.1145/3746027.3754739}
\acmISBN{979-8-4007-2035-2/2025/10}

\acmSubmissionID{376}

\begin{document}

\title{\modelname: Accelerating Lossless Compression via Lightweight Probability Models and Decoupled Parallel Dataflow}

% CCS Concepts



%Keywords
% Keywords. The author(s) should pick words that accurately describe
% the work being presented. Separate the keywords with commas.


% A "teaser" image appears between the author and affiliation
% information and the body of the document, and typically spans the
% page.



% \affiliation{%
\author{Zeyi Lu}
\authornote{Equal contribution.}
\email{luzy23@mails.tsinghua.edu.cn}
\affiliation{%
  \institution{Tsinghua University}
  \city{Shenzhen}
  \country{China}
}

\author{Xiaoxiao Ma}
\authornotemark[1]
\email{maxx24@mails.tsinghua.edu.cn}
\affiliation{%
  \institution{Tsinghua University}
  \city{Shenzhen}
  \country{China}
}

\author{Yujun Huang}
\authornotemark[1]
\email{huangyj22@mails.tsinghua.edu.cn}
\affiliation{%
  \institution{Tsinghua University}
  \city{Shenzhen}
  \country{China}
}

\author{Minxiao Chen}
\email{chenminxiao@bupt.edu.cn}
\affiliation{%
  \institution{Beijing University of Posts and Telecommunications}
  \city{Beijing}
  \country{China}
}

\author{Bin Chen}
\authornote{Corresponding author.}
\email{chenbin2021@hit.edu.cn}
\affiliation{%
  \institution{Harbin Institute of
Technology, Shenzhen}
  \city{Shenzhen}
  \country{China}
}

\author{Baoyi An}
\email{anbaoyi@huawei.com}
\affiliation{%
  \institution{Huawei Technologies Ltd.}
  \city{Shenzhen}
  \country{China}
}

\author{Shu-Tao Xia}
\email{xiast@sz.tsinghua.edu.cn}
\affiliation{%
  \institution{Tsinghua University}
  \city{Shenzhen}
  \country{China}
}

\begin{abstract}
% 1.背景&任务介绍（1句话）
The explosive growth of multi-source multimedia data has significantly increased the demands for transmission and storage, placing substantial pressure on bandwidth and storage infrastructures.
% 2.现状和问题（1句话）
While Autoregressive Compression Models (ACMs) have markedly improved compression efficiency through probabilistic prediction, current approaches remain constrained by two critical limitations: suboptimal compression ratios due to insufficient fine-grained feature extraction during probability modeling, and real-time processing bottlenecks caused by high resource consumption and low compression speeds.
% 3.本文的总体思路（1句话）
To address these challenges, we propose Efficient Dual-path Parallel Compression (EDPC), a hierarchically optimized compression framework that synergistically enhances modeling capability and execution efficiency via coordinated dual-path operations.
% 4.我们的方法（2-3句话，总分）
At the modeling level, we introduce the Information Flow Refinement (IFR) metric grounded in mutual information theory, and design a Multi-path Byte Refinement Block (MBRB) to strengthen cross-byte dependency modeling via heterogeneous feature propagation. At the system level, we develop a Latent Transformation Engine (LTE) for compact high-dimensional feature representation and a Decoupled Pipeline Compression Architecture (DPCA) to eliminate encoding-decoding latency through pipelined parallelization.
% 5.实验效果，亮点结论（1-2句话）
Experimental results demonstrate that \modelname{} achieves comprehensive improvements over state-of-the-art methods, including a 2.7× faster compression speed, and a 3.2\% higher compression ratio. These advancements establish \modelname{} as an efficient solution for real-time processing of large-scale multimedia data in bandwidth-constrained scenarios. Our code is available at \url{https://github.com/Magie0/EDPC}.
\end{abstract}

\maketitle

\section{Introduction}

\begin{figure}[t]
    \centering
    \includegraphics[width=\columnwidth]{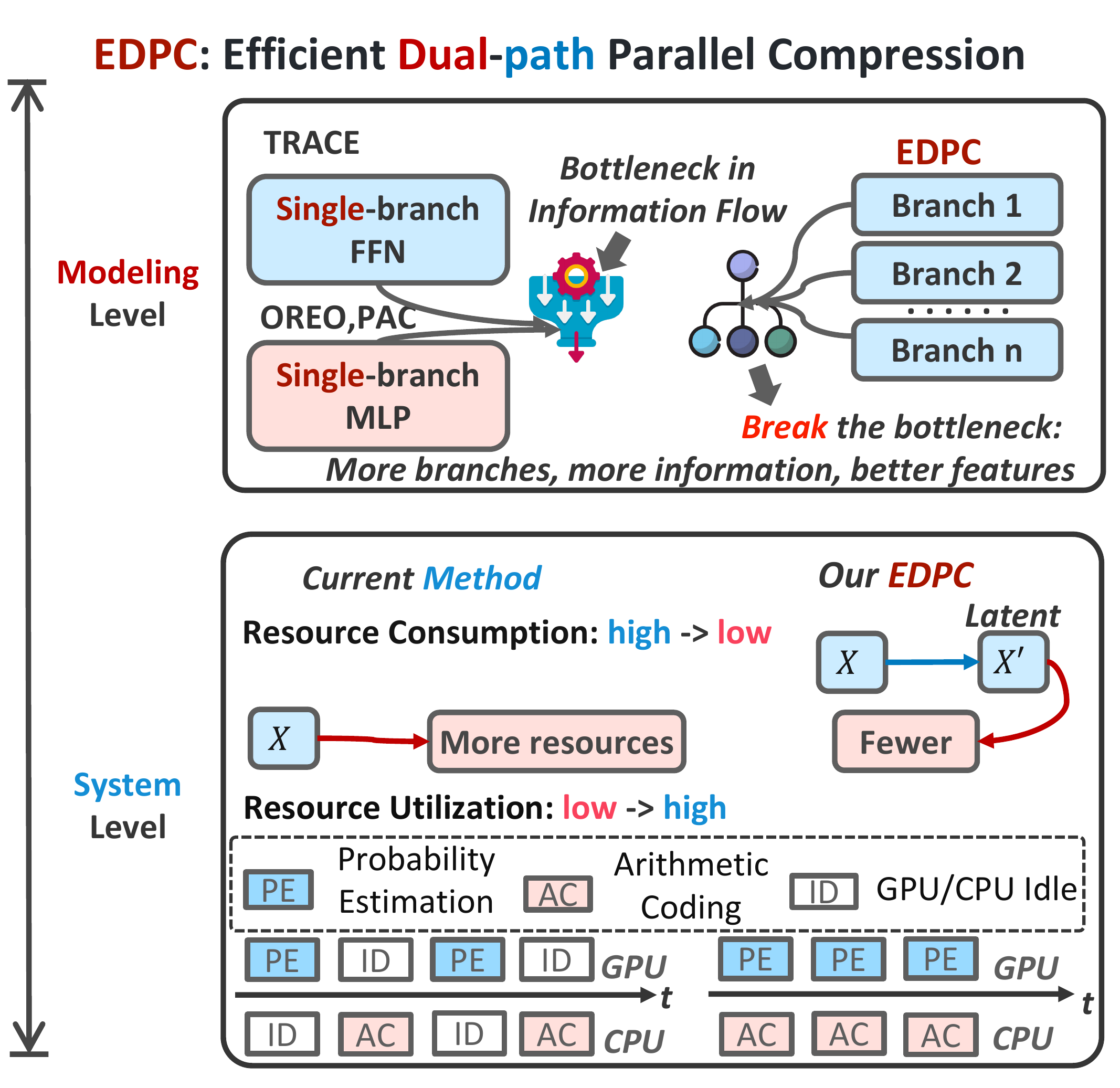}
    \vspace{-2.2em}
    \caption{We propose \textbf{EDPC}, an efficient dual-path parallel compression framework. At the \textbf{modeling level}, previous methods typically rely on single-path architectures, which restrict the diversity of information flow; in contrast, \modelname{} mitigates this bottleneck via a multi-branch design. At the \textbf{system level}, existing approaches struggle with high memory usage and underutilized hardware resources. Our framework employs latent transformation and pipelined parallelism to boost efficiency, enabling GPU-CPU collaboration.}
    \label{fig:intro}
    \vspace{-0.5cm}
\end{figure}

With the rapid development of 5G and cloud computing, the need to store and transmit massive volumes of high-resolution multimedia data in real time has become more pressing~\cite{5G,5G2,mm-videocom,MM-imagecom,RLOMM}. Traditional compression methods (e.g., Gzip~\cite{Gzip}, Zstandard~\cite{Zstd}) often fail to achieve desirable compression ratios for increasingly multimodal data. Autoregressive Compression Models (ACMs) alleviate these limitations by iteratively predicting the probability distribution of each byte conditioned on previously encoded bytes, significantly outperforming conventional methods in terms of compression ratio. Yet, current ACMs still face two major obstacles.

\textbf{First}, the \emph{information flow bottleneck} constrains representational power. Most ACMs (e.g., TRACE~\cite{DBLP:conf/www/MaoCKX22}, OREO~\cite{DBLP:conf/mm/MaoCKX22}, PAC~\cite{DBLP:conf/dac/MaoLCX23}) adopt a \emph{single-branch} design within their feature extraction modules, forcing all context information through a single path. This limits the capacity to capture diverse byte-level features and often leads to redundant representations. Although naively increasing the parameter count can improve modeling flexibility to some extent, it does not fundamentally address the single-path limitation.

\textbf{Second}, current ACMs often exhibit \emph{inefficient resource utilization}, especially for large-scale tasks such as real-time video streaming or high-resolution image transmission. For instance, NNCP~\cite{NNCP} leverages a Transformer with millions of parameters; however, its compression speed is insufficient for real-time deployment and demands excessive GPU memory. Although approaches like PAC offer partial relief, they still face significant constraints in terms of speed and memory usage. These challenges become prohibitive in scenarios that require continuous, large-scale data processing.

To tackle these issues, we propose \textbf{EDPC} (\textbf{E}fficient \textbf{D}ual-path \textbf{P}arallel \textbf{C}ompression), an integrated framework (see Figure~\ref{fig:intro}) that unifies new byte-level modeling and an efficiency system design:

\begin{itemize}
    \item \textbf{Multi-path Byte Refinement Block (MBRB).} We adopt a dual-branch feature extraction scheme to overcome the information-flow limitations of single-path architectures. To quantify how additional branches contribute unique information, we propose an \emph{Information Flow Refinement (IFR)} metric based on mutual information. Empirically, a two-branch structure offers the best trade-off between enriched feature diversity and computational overhead.
    
    \item \textbf{Latent Transformation Engine (LTE).} To drastically reduce memory and computation, we project high-dimensional features into a lower-dimensional latent space and then reconstruct them with minimal information loss. Unlike a simple dimensionality-reduction layer, LTE employs a learnable \emph{Feature Distribution Matrix} in the latent space, capturing critical inter-feature correlations. This design substantially cuts resource consumption while preserving crucial context.

    \item \textbf{Decoupled Pipeline Compression Architecture (DPCA).} Conventional ACM pipelines typically serialize probability prediction and encoding, underutilizing modern hardware. Our design decouples these stages into a concurrent pipeline, allowing GPU-based prediction and CPU-based encoding to operate in parallel. Further, we split data streams into segments for multi-process encoding, maximizing throughput by exploiting heterogeneous hardware resources.
\end{itemize}

Extensive evaluations on large-scale datasets show that \modelname{} significantly outperforms previous approaches: it reduces the parameter count by up to 4$\times$, cuts GPU memory usage nearly in half, accelerates compression speed by a factor of 2.7, and achieves a 3.2\% higher compression ratio compared to advanced baselines. In essence, \modelname{}’s multi-branch representation, latent transformation, and parallel execution work in tandem to enhance byte-level modeling while lowering system-level overhead. As a result, \modelname{} is ideally positioned for large-scale multimedia data scenarios, meeting the dual needs of high compression quality and efficient real-time processing.

\noindent \textbf{Our contributions are summarized as follows:}
\begin{enumerate}
    \item We propose an \emph{IFR} metric to measure the unique information contributed by each branch and design the \textbf{MBRB} module to enhance byte-level feature representations significantly.
    \item We introduce the \textbf{LTE} architecture, compressing features into a lower-dimensional latent space and leveraging a learnable Feature Distribution Matrix to capture context efficiently, thereby cutting memory overhead.
    \item We develop the \textbf{DPCA} pipeline, which decouples probability prediction from encoding to enable parallel GPU-CPU execution, dramatically improving compression speed.
    \item We integrate these innovations into the \textbf{EDPC} framework. Experiments show that \modelname{} not only achieves superior compression ratios but also significantly reduces GPU memory usage and encoding latency, offering an effective solution for large-scale, real-time data compression.
\end{enumerate}

\section{Related Work}

\subsection{Autoregressive Compression Model (ACM)}

\textbf{Probability Estimation.} In the current ACM, given a byte stream \(X = \{x_1, x_2, \ldots, x_{len}\}\), the model initially encodes the first \(t\) bytes with equal probability. Then, starting from the \((t+1)\)-th byte, the model uses the previous \(t\) bytes as input and calculates the conditional probability distribution for the \(i\)-th byte \(x_i\) through a forward pass~\cite{LLMZip,Tensorflow,GPT-based,LC-LLM}, 
\vspace{-0.1cm}
\begin{equation}
\Pr(i) = P(x_i \mid x_{i-t}, \ldots, x_{i-1})
\end{equation}
This process continues iteratively until all bytes are processed. At each step, the cross-entropy loss is calculated for the prediction
\vspace{-0.1cm}
\begin{equation}
\text{loss} = \text{CrossEntropy}(\Pr(i), x_i)
\end{equation}
and gradients are computed via backpropagation, followed by updates to the model parameters\cite{LLM,10889184,Dynamic-Compression,Shannon1948}.

\textbf{Data Encoding.} Once the model generates the conditional probability distribution \(\Pr(i)\), this distribution is used to encode the current byte \(x_i\)\cite{ENC}. In \modelname, we use Arithmetic Coding\cite{DBLP:conf/mm/MaoCKX22,DBLP:conf/dac/MaoLCX23,DBLP:conf/www/MaoCKX22,LC-image}.

\textbf{Overall Process.} The complete encoding and decoding procedure is outlined in Algorithm\textcolor{red}{~\ref{alg:i_byte_prediction}}. 
In the current ACM, no pre-training is required, as all operations start from randomly initialized parameters.
In line 5 of the algorithm, the ACM model estimates \(\Pr(i)\), and in line 6, \(x_i\) is encoded using arithmetic coding based on \(\Pr(i)\).
The data compression process is thus aligned with the model training process; as shown in lines 10-12 of the algorithm, parameters are updated using cross-entropy loss.
Decoding also starts with the same initial parameters, where the model computes the probability \(\Pr(i)\). Based on \(\Pr(i)\), the next byte is decoded and subsequently used as input for the next iteration.
This design eliminates the need for additional storage to retain pre-trained parameters or models.

\begin{algorithm}
\caption{Autoregressive Compression Model with Encoding and Decoding Steps}
\label{alg:i_byte_prediction}
\KwIn{Byte stream \(X = \{x_1, x_2, \ldots, x_{len}\}\), where \(len\) is the length of the input stream}
\KwOut{Compressed or decompressed byte stream}

Encode/Decode the first \(t\) bytes \(x_1, x_2, \ldots, x_t\) with equal probability\;
Initialize reconstructed bytes \(X = \{x_1, x_2, \ldots, x_t\}\)\;

\(i \leftarrow t + 1\)\;
\While{\(i \leq len\)}{
    \textbf{Encoding Step:}
    \quad \(P(x_i \mid x_{i-t}, \ldots, x_{i-1}) = \mathrm{Model}(x_{i-t}, \ldots, x_{i-1})\)\;
    \quad Encode \(x_i\) with \(P(x_i \mid x_{i-t}, \ldots, x_{i-1})\)\;

    \textbf{Decoding Step:}
    \quad \(P(x_i \mid x_{i-t}, \ldots, x_{i-1}) = \mathrm{Model}(x_{i-t}, \ldots, x_{i-1})\)\;
    \quad Decode \(x_i\) using \(P(x_i \mid x_{i-t}, \ldots, x_{i-1})\)\;
    \quad Append \(x_i\) to \(X\)\;

    \(loss = \mathrm{CrossEntropy}(P(x_i \mid x_{i-t}, \ldots, x_{i-1}), x_i)\)\;
    \(loss.backward()\)\;
    Update model parameters\;

    \(i \leftarrow i + 1\)\;
}
Output the compressed or decompressed byte stream\;

\end{algorithm}

\subsection{Mutual Information (MI)}

Mutual Information (MI) is a key concept in information theory, widely used to assess the dependency and information flow between two random variables~\cite{MI1,MI2}. In neural networks, MI is used to quantify the flow of information between different layers and evaluate its effectiveness. By analyzing MI, we can reveal the relationships between different layers of a network and subsequently optimize the network structure and training strategy.

Mutual information measures the amount of shared information between two random variables. Specifically, the mutual information \( I(X; Y) \) between two random variables \( X \) and \( Y \) is defined as:
\begin{equation}
I(X; Y) = H(X) + H(Y) - H(X, Y)
\end{equation}
where \( I(X; Y) \) represents the mutual information between random variables \( X \) and \( Y \), \( H(X) \) and \( H(Y) \) are the entropies of \( X \) and \( Y \), representing their respective uncertainties or information content, and \( H(X, Y) \) is the joint entropy of \( X \) and \( Y \), which represents the overall uncertainty when both \( X \) and \( Y \) are considered together~\cite{MI3}.

\section{Method}

\begin{figure}[t]
    \centering
    \includegraphics[width=\columnwidth]{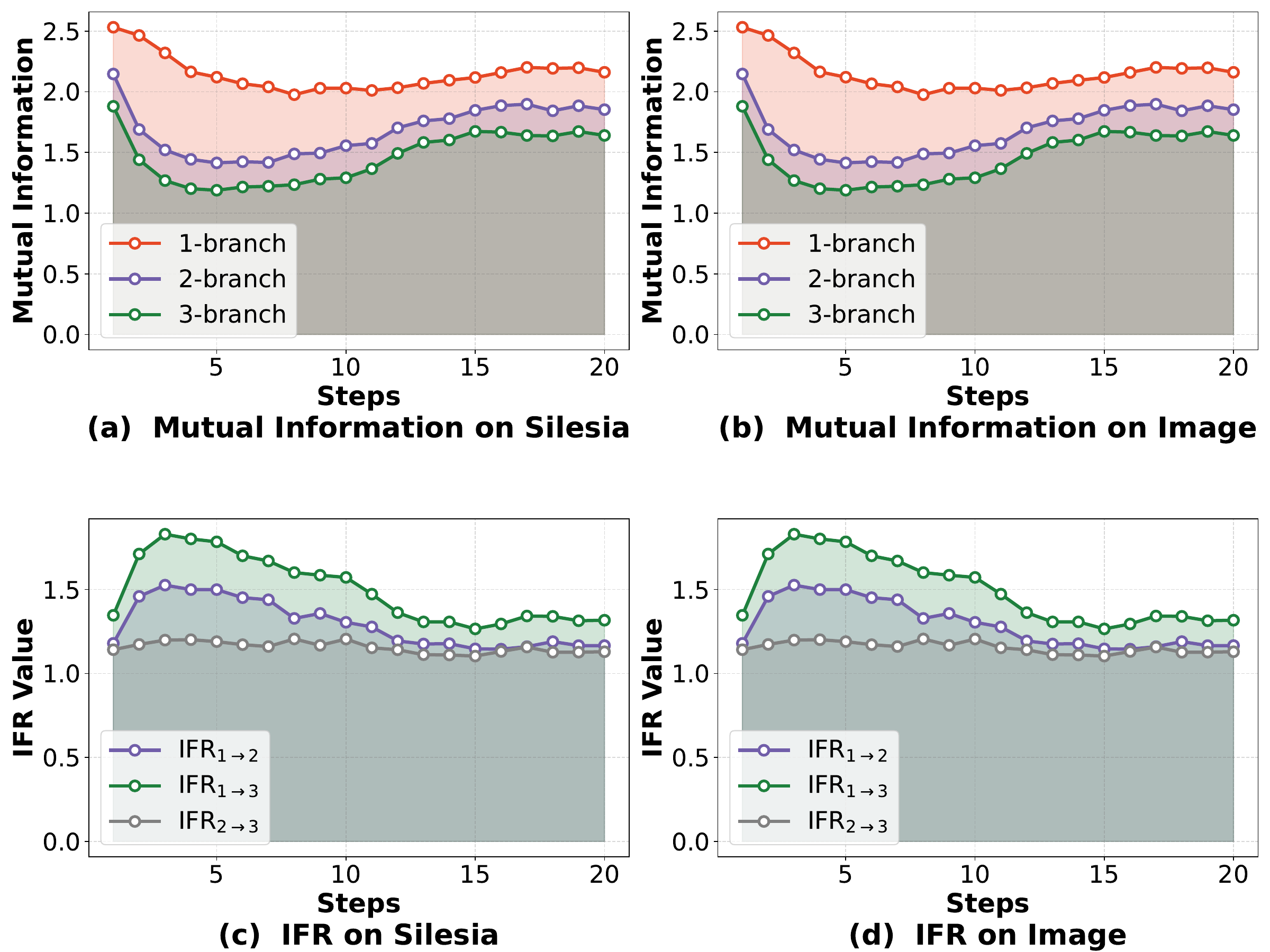}
    %\vspace{-2em}
    \caption{
Comparison of mutual information and Information Flow Refinement (IFR) across different architectures and datasets. Subfigures (a) and (b) show the mutual information between the skip connection $S$ and the fused output $S+X$, where lower mutual information indicates that the added feature $X$ contributes more diverse information beyond the residual path. Subfigures (c) and (d) depict IFR metrics, quantifying the degree of information enhancement brought by multi-branch structures. Higher IFR values suggest richer information flow and more effective feature refinement.
}
    \label{fig:pre_exp}
    \vspace{-0.4cm}
\end{figure}

\begin{figure*}[t]
    \centering
    \includegraphics[width=\textwidth]{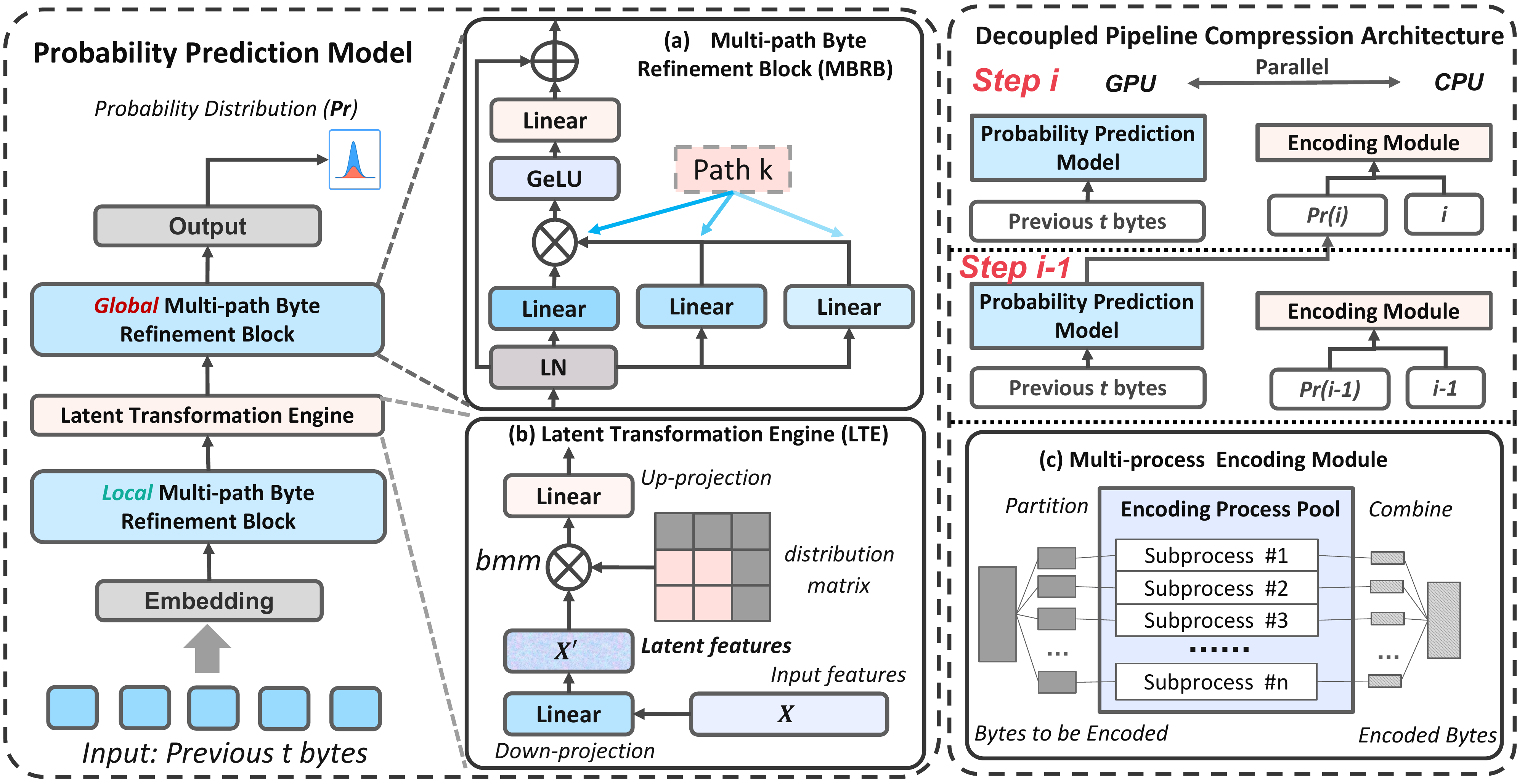}
    \vspace{-2em}
\caption{
Overview of \modelname{} (best viewed in color). 
The left part shows the Probability Prediction Model, which efficiently predicts the probability distribution of the next byte. It consists of two Multi-path Byte Refinement Blocks (MBRB) and a Latent Transformation Engine (LTE). 
The MBRB (a) adopts a multi-branch architecture to enhance byte-level feature extraction, while the LTE (b) projects features into a latent space to reduce computational and memory cost. 
The right part depicts the parallel compression architecture. 
The proposed Decoupled Pipeline Compression Architecture (DPCA) enables concurrent GPU prediction (step $i$) and CPU encoding (step $i{-}1$), improving overall efficiency. 
A multi-process encoding module (c) further accelerates encoding by partitioning the stream into parallel subprocesses.}
    \label{fig:arc}
    \vspace{-0.4cm}
\end{figure*}

\subsection{Multi-Path Byte Refinement Block (MBRB)}

Contemporary compression architectures—such as the FFN in TRACE and the MLP modules in OREO and PAC—predominantly employ single-branch structures, wherein information flows along a single pathway. While simple and efficient, such designs inherently limit the diversity of information propagation, thus constraining the model's capacity to extract rich and complementary features from byte sequences. This reliance on a single path can hinder the network’s ability to capture complex correlations and feature interactions, especially in scenarios involving multi-source data. To address these limitations, we begin by analyzing information flow through the lens of mutual information, and on this basis, propose the \textit{Multi-Path Byte Refinement Block (MBRB)}.

\subsubsection{\textbf{Information Flow Refinement (IFR)}}
From the perspective of Mutual Information (MI)—a measure that quantifies the dependency and information exchange between random variables—single-branch architectures impose constraints on the diversity and redundancy of extracted information. In contrast, multi-branch architectures enable data to be processed through multiple parallel feature pathways, thereby enhancing the diversity of information flow, which can lead to more robust feature representations.

In a single-branch architecture, let \( S \in \mathbb{R}^{b \times d} \) denote the skip connection and \( X \in \mathbb{R}^{b \times d} \) the extracted feature. The fused output is given by \( S + X \in \mathbb{R}^{b \times d} \). The corresponding mutual information is:
\begin{equation}
I_{SB}(S; S + X) = H(S) + H(S + X) - H(S, S + X)
\end{equation}
where \( H(S) \) and \( H(S+X) \) are the entropies of the skip connection and the fused output, respectively, and \( H(S, S + X) \) is their joint entropy.

For a general multi-branch architecture with \( k \) branches, we define the aggregated feature as \( X^{(k)} = \sum_{i=1}^{k} X_i \in \mathbb{R}^{b \times d} \), where each \( X_i \) represents the output of the \( i \)-th branch. The fused output becomes \( S + X^{(k)} \). The mutual information in this context is:
\begin{equation}
I_{MB}^{(k)}(S; S + X^{(k)}) = H(S) + H(S + X^{(k)}) - H(S, S + X^{(k)})
\end{equation}

To unify the notation across both single- and multi-branch settings, we define:
\begin{equation}
I^{(k)}(S; S + X^{(k)}) =
\begin{cases}
I_{SB}(S; S + X), & \text{if } k = 1 \\
I_{MB}^{(k)}(S; S + X^{(k)}), & \text{if } k > 1
\end{cases}
\end{equation}

To estimate the mutual information, we adopt Kraskov’s non-parametric estimator~\cite{Kraskov_2004,Kraskov_2,Kraskov_3}:
\begin{equation}
I(Z; Y) \approx \psi(v) + \psi(c) - \frac{1}{c} \sum_{i=1}^{c} \left( \psi(c_z^{(i)} + 1) + \psi(c_y^{(i)} + 1) \right)
\end{equation}
where \( \psi(\cdot) \) is the Digamma function, \( v \) is the number of nearest neighbors, \( c \) denotes the number of samples, and \( c_z^{(i)} \), \( c_y^{(i)} \) represent the number of neighbors for sample points \( z_i \) and \( y_i \), respectively.

To quantify the enhancement in information flow facilitated by increasing the number of branches, we propose a generalized metric termed \emph{Information Flow Refinement (IFR)}. IFR evaluates the relative improvement in information propagation and redundancy control of a \( k_2 \)-branch structure over a \( k_1 \)-branch structure and is defined as:
\begin{equation}
\text{IFR}_{k_1 \rightarrow k_2} = \frac{I^{(k_1)}(S; S + X^{(k_1)})}{I^{(k_2)}(S; S + X^{(k_2)})}
\end{equation}

As shown in Figure~\ref{fig:pre_exp}(a) and (b), the mutual information decreases as the number of branches increases, indicating that multi-branch structures are able to introduce more novel byte-level features beyond the residual pathway. This suggests that additional branches can enhance the diversity of the extracted information and reduce redundancy. Figure~\ref{fig:pre_exp}(c) and (d) further demonstrate that both two-branch (IFR$_{1 \rightarrow 2}$) and three-branch (IFR$_{1 \rightarrow 3}$) architectures exhibit higher IFR values compared to the single-branch baseline, implying that the additional branches effectively contribute more independent and complementary information. In particular, the three-branch architecture consistently achieves the highest IFR values, highlighting its superiority in promoting efficient information flow and reducing redundancy.

\vspace{-0.2cm}
\subsubsection{\textbf{Multi-Path Byte Refinement Block (MBRB)}}

While a three-branch structure surpasses both one- and two-branch designs in information flow, the incremental benefit from adding a third branch is notably smaller than the jump from one to two branches. This diminishing-return effect indicates that the second branch captures most of the beneficial complementary information, leaving the third branch to offer only marginal improvements (IFR\(_{2 \rightarrow 3}\)). Thus, a two-branch structure strikes an ideal balance between performance and overhead, making it a practical default choice for compression-oriented architectures.

Motivated by this insight, we propose the \textbf{Multi-Path Byte Refinement Block (MBRB)}, a compact and effective module designed to enhance byte-level representation learning in lossless compression. MBRB adopts a two-branch structure to increase the diversity of information flow while maintaining computational efficiency. Unlike TRACE that rely on self-attention for context modeling, MBRB retains the MLP-based design due to its architectural simplicity and proven effectiveness in byte-level modeling~\cite{DBLP:conf/mm/MaoCKX22,DBLP:conf/dac/MaoLCX23,gateMLP}. It introduces multiple parallel transformation paths, whose outputs are fused through element-wise interaction.

Let the input be \( X \in \mathbb{R}^{b \times F} \), where \( b \) is the batch size and \( F \) is the feature dimension. The MBRB block operates as follows:
\vspace{-0.1cm}
\begin{align}
X_0 &= \text{LN}(X) \\
X_i &= \text{Linear}_i(X_0), \quad i = 1, 2, \dots, k \\
X_{\text{fused}} &= X_1 \odot X_2 \odot \cdots \odot X_k \\
X_{\text{ff}} &= \text{GeLU}(X_{\text{fused}}) \\
X_{\text{out}} &= \text{Linear}_{\text{out}}(X_{\text{ff}}) + X
\end{align}
\vspace{-0.5cm}

Here, \( \odot \) denotes element-wise multiplication, \( \text{LN}(\cdot) \) is layer normalization, and each \( \text{Linear}_i(\cdot) \) represents a learnable transformation path. This multiplicative fusion encourages the model to focus on consistent and shared feature dimensions across branches. Finally, a residual connection is added to preserve input fidelity and support stable gradient flow.

\vspace{-0.1cm}
\subsection{Latent Transformation Engine (LTE)}

In real-world high-dimensional data processing tasks, traditional computational methods often suffer from excessive memory consumption and poor computational efficiency. These challenges become increasingly pronounced as both the feature dimension and batch size scale up. To address this bottleneck, particularly in large-scale scenarios, we propose a novel architectural module called the \textbf{Latent Transformation Engine (LTE)}. The core idea of LTE is to project input features into a lower-dimensional latent space through a compression–recovery mechanism, thereby significantly reducing computational cost and memory usage while preserving the model's expressive capacity.

LTE consists of two primary stages: a down-projection (compression) phase and an up-projection (recovery) phase.

In the down-projection phase, the input feature matrix \( X \in \mathbb{R}^{b \times F} \), where \( b \) denotes the batch size and \( F \) the original feature dimension, is linearly mapped to a lower-dimensional latent space. Given a predefined compression ratio \( r \), the reduced feature dimension is \( F' = F / r \). The transformation is expressed as:
\vspace{-0.1cm}
\begin{equation}
X' = W_1 X + b_1, \quad X' \in \mathbb{R}^{b \times F'}
\end{equation}
\vspace{-0.1cm}
where \( W_1 \in \mathbb{R}^{F' \times F} \) is the weight matrix for the down-projection, and \( b_1 \) is the bias term.

To enhance the modeling of feature-wise interactions in this latent space, LTE introduces a learnable three-dimensional tensor \( \mathcal{U} \in \mathbb{R}^{b \times F' \times F'} \), referred to as the \textbf{Feature Distribution Matrix (FDM)}. This matrix is initialized randomly and updated during training. It serves to dynamically encode the structural correlations and distributional patterns between latent features for each instance.
The compressed features \( X' \) are then processed via batch matrix multiplication with the FDM:
\vspace{-0.1cm}
\begin{equation}
Y = \text{bmm}(X', \mathcal{U}), \quad Y \in \mathbb{R}^{ b \times F'}
\end{equation}
This operation efficiently captures contextual dependencies across feature dimensions while operating in a compact latent space, thereby improving scalability and runtime efficiency.

To recover the original feature representation, the second stage of LTE performs an up-projection that maps the latent representation back to the original high-dimensional space:
\vspace{-0.1cm}
\begin{equation}
X'' = W_2 Y + b_2, \quad X'' \in \mathbb{R}^{ b \times F}
\end{equation}
where \( W_2 \in \mathbb{R}^{F \times F'} \) and \( b_2 \) are the up-projection weight and bias, respectively.
The dual-stage compression–recovery design of LTE mitigates the resource bottleneck associated with high-dimensional modeling, while enabling expressive and adaptive feature transformation. 
%This makes LTE particularly well-suited for the lossless compression of large-scale multimedia data, where precise feature representation and resource efficiency are both critical requirements.

\subsection{Decoupled Pipeline Compression Architecture (DPCA)}

In traditional ACM architectures, the compression process follows a strict sequential execution model, where probability prediction for time step \( t \) must be completed before encoding for step \( t \) can begin. This sequential execution leads to significant resource wastage, with approximately 50\% of hardware resources remaining idle, severely limiting compression speed. Additionally, this architecture does not fully leverage the parallel computation capabilities of modern hardware, resulting in inefficiencies and suboptimal resource utilization, which hinder the potential for performance improvements.

To address this issue, we propose the \textbf{Decoupled Pipeline Compression Architecture (DPCA)}, a new compression system design. The core innovation of DPCA lies in decoupling the probability prediction and encoding processes, eliminating their time dependency. Specifically, DPCA utilizes a pipelined design, allowing the probability prediction and encoding processes to run in parallel. While probability prediction for step \( t+1 \) is being performed, encoding for step \( t \) can proceed concurrently. This significantly improves data throughput and reduces overall compression time.

Furthermore, DPCA fully exploits the synergy between GPU and CPU to enhance hardware resource utilization while preventing idle resource waste. By overlapping slower GPU processing with faster CPU encoding tasks, DPCA optimally utilizes the computational power of both the CPU and GPU, leading to a substantial speedup.

\begin{figure}[t]
    \centering
    \includegraphics[width=\columnwidth]{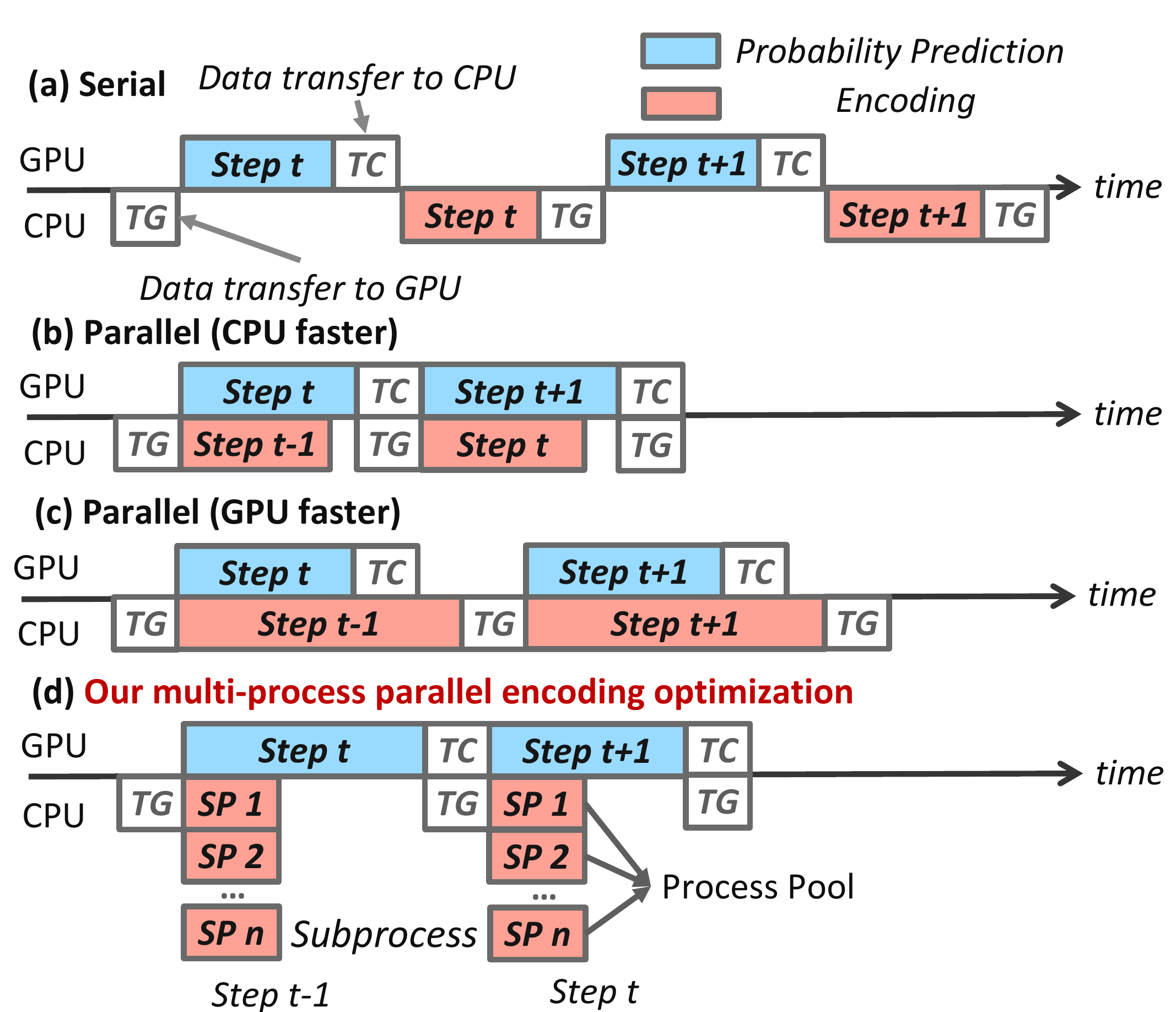}
    \vspace{-2em}
    \caption{Comparison of different encoding strategies: (a) serial processing, (b) parallel processing with faster CPU, (c) parallel processing with faster GPU, and (d) our proposed multi-process parallel encoding optimization.}
    \label{fig:par}
    \vspace{-0.5cm}
\end{figure}

To further improve compression efficiency, DPCA introduces \textbf{multi-process parallel encoding optimization} (as shown in Fig.~\ref{fig:par}(d)). The byte stream of each batch is divided into multiple independent segments, which are then encoded in parallel by subprocesses within a process pool. The number of subprocesses can be dynamically adjusted based on the CPU capabilities of the underlying hardware, ensuring optimal performance across various platforms. This parallel encoding strategy significantly reduces encoding time and maximizes resource efficiency~\cite{FSE}.

In conclusion, DPCA breaks the bottleneck of traditional ACM architectures through decoupling and pipelined design, enabling the compression process to run efficiently in a parallel environment. Additionally, by incorporating multi-process parallel encoding and GPU-CPU collaboration, DPCA optimizes hardware resource usage, drastically reduces compression time, and drives a revolutionary breakthrough in lossless compression speed. The innovative design of DPCA has made significant advancements in the lossless compression field, particularly in improving compression efficiency and speed, achieving unprecedented breakthroughs.

\section{Experiment}

\subsection{Experimental Setup}

Experiments are benchmarked on datasets sourced from various domains~\cite{Lossless-FCBench,Lossless-Floating-Point,Lossless-ALP}. A diverse range of real-world datasets with different data types are considered, with detailed descriptions provided in Table ~\ref{tab:datasets}. We use compression ratio, compression speed, GPU memory usage, and the total number of parameters as evaluation metrics. All experiments are conducted on a GPU with 34.10 TFLOPS FP32 performance and equipped with 12GB of high-bandwidth memory. The experiments were performed using this GPU in conjunction with an Intel(R) Xeon(R) Platinum 8269CY @ 2.50GHz CPU. To ensure fairness, all ACM methods follow the dynamic compression procedure, with no pretraining involved\cite{DBLP:conf/mm/MaoCKX22,DBLP:conf/dac/MaoLCX23,DBLP:conf/www/MaoCKX22,DBLP:conf/dcc/GoyalTCO21,NNCP,DBLP:conf/dcc/GoyalTCO21,ma2025msdzip}.

In our experiments, the number of historical symbols is fixed at 16, and other settings follow the standard configurations described in the corresponding papers. In the proposed \modelname{}, the Local and Global MBRBs employ MLPs with hidden layer dimensions of 2048 and 4096, respectively. Additionally, the Latent Transformation Engine (LTE) is configured with a compression ratio of \( r = 4 \). In the encoding module, 32 subprocesses are used in the thread pool, resulting in a chunk size of 32. The model is optimized using the Adam optimizer with a learning rate of 0.001~\cite{Adam,Lossless-Chimp,LosslessCodeCompression}.

\begin{table}[t]
  \caption{Description of Compression Datasets}
  \label{tab:datasets}
  \setlength{\tabcolsep}{3pt}
  \renewcommand{\arraystretch}{1.05}
  \small
  %\footnotesize
  \centering
  \begin{tabular}{lll}
    \toprule
    \textbf{Name} & \textbf{Size} & \textbf{Description} \\
    \midrule
    Book     & 1000 MB & First 1000M bytes of BookCorpus~\cite{DBLP:conf/nips/VaswaniSPUJGKP17}. \\
    Enwik9   & 1000 MB & First 1000M bytes of English Wikipedia~\cite{Mahoney2006}. \\
    Float    & 1.1 GB  & Spitzer Space Telescope data~\cite{DBLP:journals/tc/BurtscherR09}. \\
    Sound    & 842 MB  & ESC~\cite{DBLP:conf/mm/Piczak15} dataset for environmental sound. \\
    Image    & 1.2 GB  & 100{,}000 images from ImageNet~\cite{DBLP:conf/cvpr/DengDSLL009}. \\
    Backup   & 1000 MB & Random 1000M byte extract from a disk backup. \\
    Silesia  & 206 MB  & A standard compression benchmark~\cite{Deorowicz1985}. \\
    \bottomrule
  \end{tabular}
  \vspace{-0.5cm}
\end{table}

\subsection{Compression Ratio Evaluation}

As shown in Table~\ref{tab:compression_ratios}, the compression ratios of various methods are evaluated across a comprehensive collection of large-scale datasets, encompassing both homogeneous and heterogeneous data types. Traditional methods such as Gzip~\cite{Gzip}, 7z~\cite{7z}, and Zstd-19~\cite{Zstd} show limited performance, particularly on complex data like \textit{Image} and \textit{Float}. Deep learning-based approaches outperform traditional ones by a substantial margin, with PAC achieving the strongest results among prior models. \modelname{} further improves upon these results and consistently achieves the best compression ratios across all datasets. Compared to PAC, it yields relative gains of 7.71\% on \textit{Enwik9}, 3.56\% on \textit{Book}, 1.33\% on \textit{Sound}, and 2.33\% on \textit{Float}. Notably, even on more heterogeneous datasets such as \textit{Silesia} and \textit{Backup}, \modelname{} attains improvements of 6.41\% and 1.04\%, respectively. These results highlight \modelname{}'s robustness and versatility in handling both structured and diverse real-world data distributions.

In terms of average relative improvements, \modelname{} delivers consistent gains over all baselines. Compared to traditional methods, it improves compression ratios by an average of 68.95\% over Gzip, 33.71\% over 7z, and 45.73\% over Zstd-19. Among deep learning-based approaches, \modelname{} surpasses Dzip, TRACE, and OREO by 18.12\%, 10.91\%, and 5.99\%, respectively. Even against the strongest prior method PAC, \modelname{} achieves an overall gain of 3.20\%, with a 2.99\% average improvement on homogeneous data and 3.73\% on heterogeneous data. These results collectively demonstrate \modelname{}'s superior capability to generalize across both structured and complex, diverse data distributions.

\begin{table*}[t]
\centering
\caption{Compression Ratios on Large Datasets. \textbf{\modelname{}} consistently outperforms both traditional and deep learning-based methods. We report the average relative improvement (\%) of \textbf{\modelname{}} over each method in homogeneous, heterogeneous, and overall settings.}
\label{tab:compression_ratios}
\setlength{\tabcolsep}{6pt}
\begin{tabular}{lccccc c ccc c}
\toprule
\multirow{2}{*}{\textbf{Methods}} & \multicolumn{5}{c}{\textbf{Homogeneous Data}} &
\makecell{\textbf{Hom.}\\\textbf{Gain (\%)}}& \multicolumn{2}{c}{\textbf{Heterogeneous Data}} & \makecell{\textbf{Het.}\\\textbf{Gain (\%)}} & \makecell{\textbf{Overall}\\\textbf{Gain (\%)}} \\
\cmidrule(lr){2-6} \cmidrule(lr){8-9}
 & \textit{Enwik9} & \textit{Book} & \textit{Sound} & \textit{Image} & \textit{Float} & & \textit{Silesia} & \textit{Backup} & & \\
\midrule
\multicolumn{11}{l}{\textit{Traditional Methods}} \\
Gzip     & 3.09 & 2.77 & 1.37 & 1.14 & 1.06 & 71.96 & 3.10 & 1.28 & \multicolumn{1}{c}{61.43} & \multicolumn{1}{c}{68.95} \\
7z       & 4.35 & 3.80 & 1.59 & 1.38 & 1.14 & 37.33 & 4.25 & 1.56 & \multicolumn{1}{c}{24.65} & \multicolumn{1}{c}{33.71} \\
Zstd-19  & 4.24 & 3.73 & 1.40 & 1.16 & 1.10 & 48.74 & 3.97 & 1.36 & \multicolumn{1}{c}{38.20} & \multicolumn{1}{c}{45.73} \\
\midrule
\multicolumn{11}{l}{\textit{Deep Learning-based Methods}} \\
Dzip     & 4.47 & 3.95 & 2.04 & 1.72 & 1.26 & 21.35 & 4.78 & 1.78 & \multicolumn{1}{c}{10.04} & \multicolumn{1}{c}{18.12} \\
TRACE    & 5.29 & 4.58 & 2.16 & 1.81 & 1.28 & 10.54 & 4.63 & 1.78 & \multicolumn{1}{c}{11.84} & \multicolumn{1}{c}{10.91} \\
OREO     & 5.68 & 4.94 & 2.25 & 1.86 & 1.28 & 5.78  & 4.86 & 1.87 & \multicolumn{1}{c}{6.50}  & \multicolumn{1}{c}{5.99} \\
PAC      & 5.97 & 5.05 & 2.25 & 1.96 & 1.29 & 2.99  & 4.99 & 1.92 & \multicolumn{1}{c}{3.73}  & \multicolumn{1}{c}{3.20} \\
\rowcolor{pink!40}
\textbf{\modelname{}} & \textbf{6.43} & \textbf{5.23} & \textbf{2.28} & \textbf{1.96} & \textbf{1.32} & \textbf{0} & \textbf{5.31} & \textbf{1.94} & \multicolumn{1}{c}{\textbf{0}} & \multicolumn{1}{c}{\textbf{0}} \\
\bottomrule
\end{tabular}
\vspace{-6pt}
\end{table*}

\begin{table}[htbp]
\centering
\small
\setlength{\tabcolsep}{6pt}
\renewcommand{\arraystretch}{1.15}
\caption{Compression Speed Comparison of Traditional and Neural Compressors.}
\label{tab:ar_compressor_speed}
\begin{tabular}{llcc}
\toprule
\textbf{Type} & \textbf{Compressor} & \textbf{Speed} & \textbf{Rel. Gain (\%)} \\
\midrule
Traditional & Gzip     & 983.46 MB/min     & --     \\
Traditional & 7z    & 1966.98 MB/min     & --     \\
Traditional & Zstd-19     & 120.34 MB/min     & --     \\
\midrule
Neural      & NNCP     & 261$\sim$432.6 KB/min & $>2260$ \\
Neural      & Dzip     & 645.6 KB/min  & 1482.6 \\
Neural      & TRACE    & 1422.6 KB/min  & 618.0  \\
Neural      & PAC      & 3789.6 KB/min  & 169.5  \\
\rowcolor{pink!20}
Neural      & \textbf{\modelname{}}    & \textbf{10213.8 KB/min} & \textbf{0} \\
\bottomrule
\end{tabular}
\vspace{-0.5cm}
\end{table}

\subsection{Efficiency Evaluation}

We assess the efficiency of \modelname{} from two perspectives: compression speed and resource consumption.

\begin{figure}[t]
\vspace{-0cm}
    \centering
    \includegraphics[width=\columnwidth]{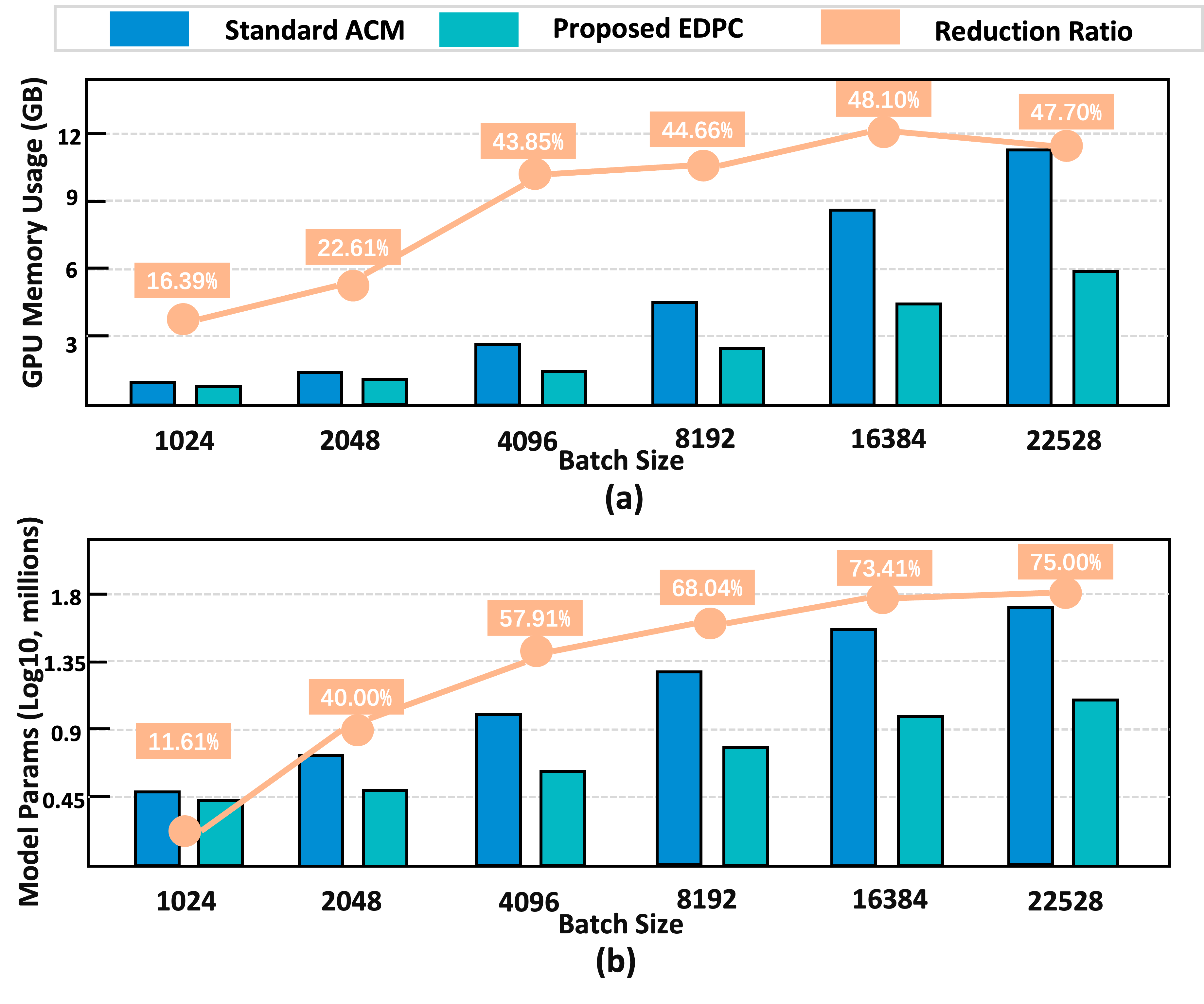}
    \setlength{\abovecaptionskip}{-0.4cm} 
\caption{GPU memory usage (a) and total model parameters (b) of the proposed \modelname{} across different batch sizes. The lines indicate the reduction ratios compared to the PAC.}
\label{fig:efficiency}
\vspace{-0.2cm}
\end{figure}

Table~\ref{tab:ar_compressor_speed} compares the compression speeds of neural-based compressors. \modelname{} significantly outperforms all baselines, achieving a throughput of 170.23~KB/s, which is 2.7$\times$ faster than PAC and over 7$\times$ faster than TRACE. Compared to the earliest method NNCP, the speed improvement exceeds 23$\times$ based on the upper bound estimate.\footnotemark[1] These results demonstrate that \modelname{} is well suited for latency-sensitive applications and large-scale deployment.
Next, we evaluate the GPU memory usage and parameter count of \modelname{} under different batch sizes, in comparison with the standard ACM baseline. As shown in Figure~\ref{fig:efficiency}(a), \modelname{} consistently achieves lower memory usage, with up to 48.1\% reduction observed at larger batch sizes. Figure~\ref{fig:efficiency}(b) further highlights \modelname{}'s parameter efficiency: the number of model parameters decreases sharply as batch size increases, reaching a maximum reduction of 75\%. These efficiency gains are primarily attributed to the lightweight architectural design and the introduction of the Latent Transformation Engine (LTE), which compresses feature representations while minimizing performance degradation.

\subsection{Proposed Method Analysis}  \label{subsec:ablation_study}

\subsubsection{\textbf{Ablation Study of \modelname{}}}

\begin{figure}[t]
\vspace{-0.2cm}
    \centering
    \includegraphics[width=\columnwidth]{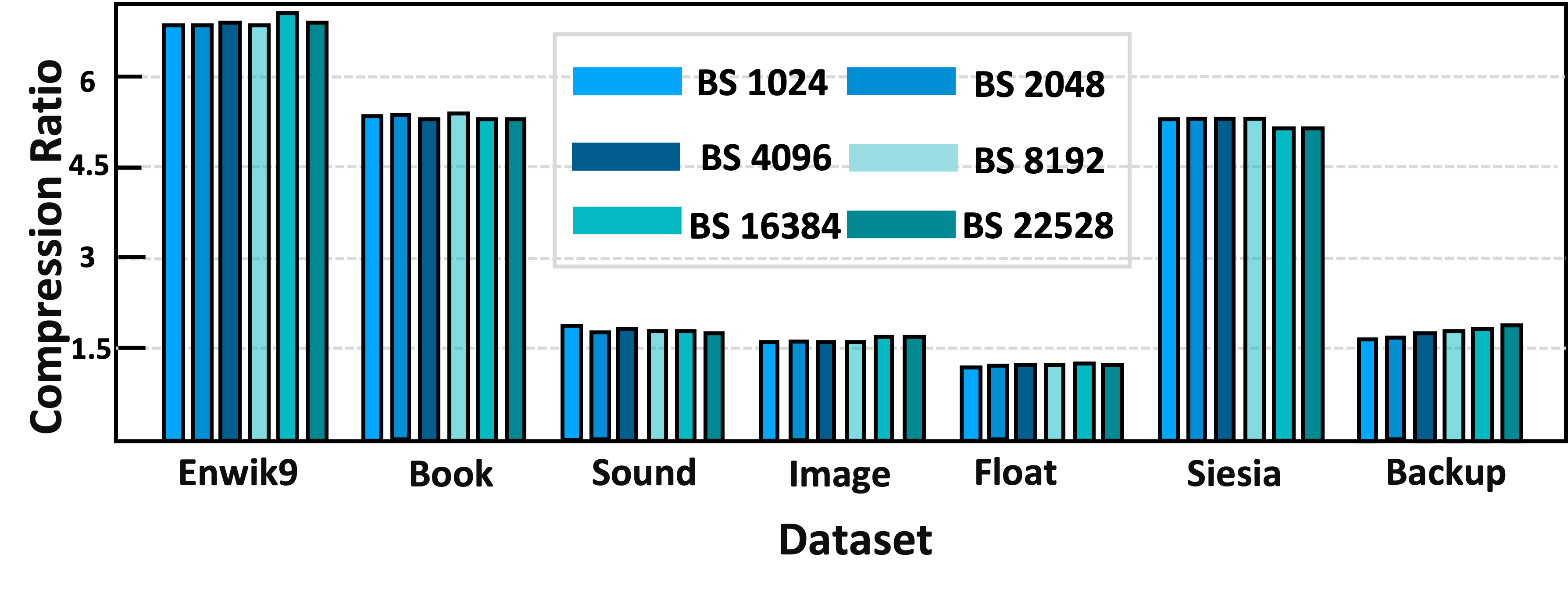}
    \setlength{\abovecaptionskip}{-0.4cm} 
\caption{Compression ratios of \modelname{} across different batch sizes.}
\label{CR}
\vspace{-0.6cm}
\end{figure}

We first evaluated the compression performance of the proposed \modelname{} across multiple datasets and a wide range of batch sizes, as shown in Figure~\ref{CR}. The results show that \modelname{} consistently achieves stable compression ratios under all settings.

Next, we conduct a comprehensive ablation study to evaluate the contribution of each proposed component in \modelname{}, as summarized in Table~\ref{tab:ablation_compression}. The results demonstrate that both the Local MBRB and Global MBRB are critical to achieving strong compression performance. The LTE plays a central role in reducing GPU memory usage and model parameter count, with its removal leading to up to 4.1$\times$ memory overhead and nearly 7$\times$ parameter increase. Notably, the pipeline architecture significantly improves compression speed, and LTE also contributes to acceleration by reducing the computational cost.

\begin{table}[t]
\centering
\caption{Ablation study of \modelname{} on the Silesia dataset (batch size = 4096). We report compression ratio, GPU memory usage, parameter count, and compression speed.}
\small
\begin{tabular}{lcccc}
\toprule
Ablation & Ratio ↑ & Mem ↓ & Params ↓ & Speed (kB/s) ↑\\
\midrule
\rowcolor{pink!40} \textbf{\modelname{}}             & 5.29 & 1546MB & 4.15e+07 & 157 \\  
\midrule\midrule
w/o LTE              & 5.33 & 6374MB & 2.93e+08 & 73  \\  
low-rank factorization & 5.09 & 5924MB & 1.76e+08 & 80 \\
\midrule\midrule
w/o Local MBRB       & 5.16 & 1358MB & 3.84e+07 & 201 \\
w/o Global MBRB      & 4.98 & 1282MB & 3.52e+07 & 258 \\
\midrule\midrule
w/o pipeline         & 5.29 & 1546MB & 4.15e+07 & 70  \\
\bottomrule
\end{tabular}
\label{tab:ablation_compression}
\vspace{-0.2cm}
\end{table}

\begin{table}[t]
\centering
\caption{Comparison of 2-branch and 3-branch MBRB designs on two datasets. While the 3-branch structure slightly improves compression ratio, it incurs higher memory and computational cost(batchsize = 8192).}
\label{tab:mbrb_comparison}
\small
\begin{tabular}{lcccc}
\toprule
\textbf{Dataset} & \textbf{Branch} & \textbf{Ratio ↑} & \textbf{Speed (kB/min) ↑} & \textbf{Mem (MB) ↓} \\
\midrule
\multirow{2}{*}{\textit{Backup}} 
& 2-branch & 1.94 & 10214.8 & 2592 \\
& 3-branch & 1.96 & 7826.4 & 2820 \\
\midrule
\multirow{2}{*}{\textit{Silesia}} 
& 2-branch & 5.31 & 10113.5 & 2592 \\
& 3-branch & 5.34 & 7817.8 & 2820 \\
\bottomrule
\end{tabular}
\end{table}

\subsubsection{\textbf{Efficiency and Trade-off Analysis of MBRB}}

To assess the impact of the number of branches in MBRB, we compare 2-branch and 3-branch variants on two representative datasets, as shown in Table~\ref{tab:mbrb_comparison}. The results show that increasing the number of branches from two to three yields only marginal improvements in compression ratio—an increase of just 0.02 on \textit{Backup} and 0.03 on \textit{Silesia}. However, this minor gain comes at a notable cost: GPU memory usage increases from 2592 MB to 2820 MB (a rise of 8.8\%), and compression speed drops by 23.4\% on \textit{Backup} and 22.7\% on \textit{Silesia}. These findings suggest that the 2-branch MBRB strikes a desirable balance between compression effectiveness and system efficiency. It achieves nearly optimal performance while maintaining high throughput and low memory consumption, making it the preferred choice for practical deployment.

\subsubsection{\textbf{Effectiveness Analysis for LTE}}

\begin{figure}[t]
    \centering
    \includegraphics[width=\columnwidth]{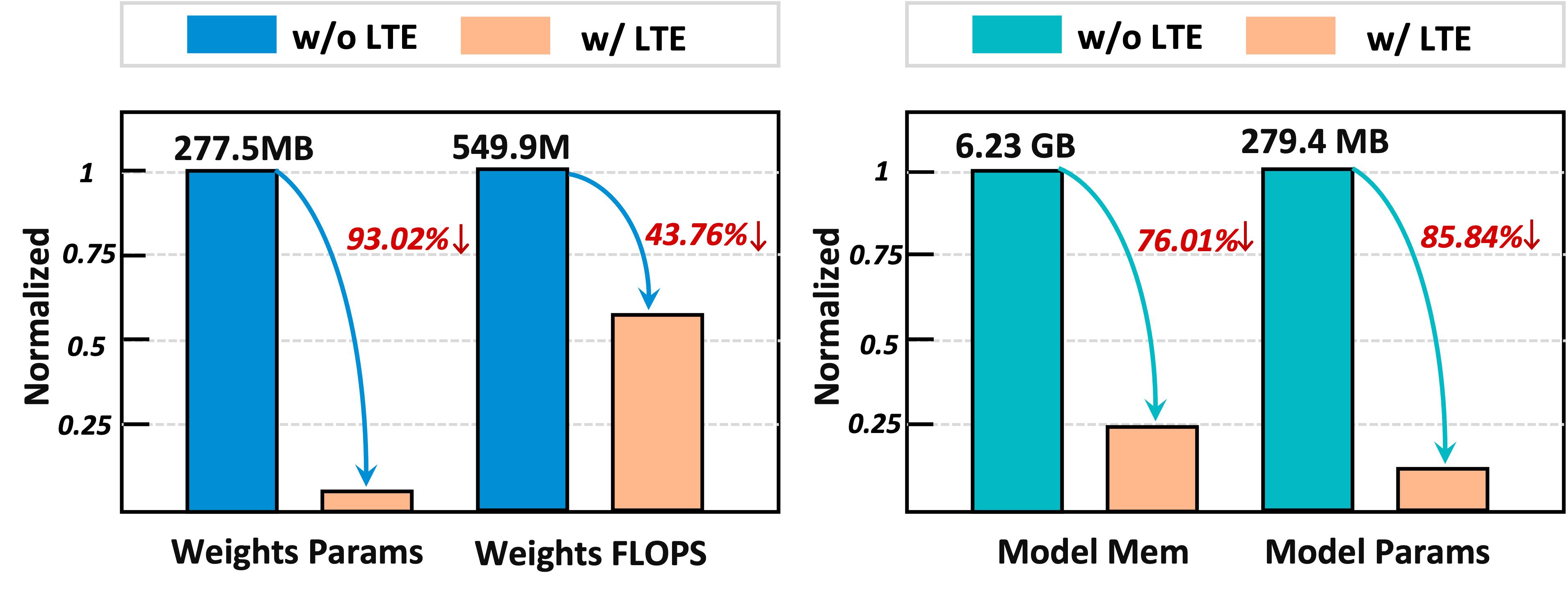}
    \setlength{\abovecaptionskip}{-0.4cm} 
\caption{Actual performance comparison of with LTE and without LTE within \modelname{} (Batchsize=4096).}
\label{LTE}
\vspace{-0.2cm}
\end{figure}

We further analyze the effectiveness of the Latent Transformation Engine (LTE) within the \modelname{} framework. As shown in Figure~\ref{LTE}, the left panel compares the parameter count and FLOPs of the Feature Distribution Matrix (FDM) module. Leveraging a dimensionality reduction design, LTE reduces the number of weight parameters by 93.02\% and lowers the computational cost (FLOPs) by 43.76\%. 
The right panel presents the overall model-level efficiency. When integrated into \modelname{}, LTE yields a 76.01\% reduction in GPU memory usage and an 85.84\% reduction in total parameters compared to the non-LTE baseline.

\subsubsection{\textbf{Effectiveness of DPCA}}

\begin{figure}[t]
    \centering
    \includegraphics[width=\columnwidth]{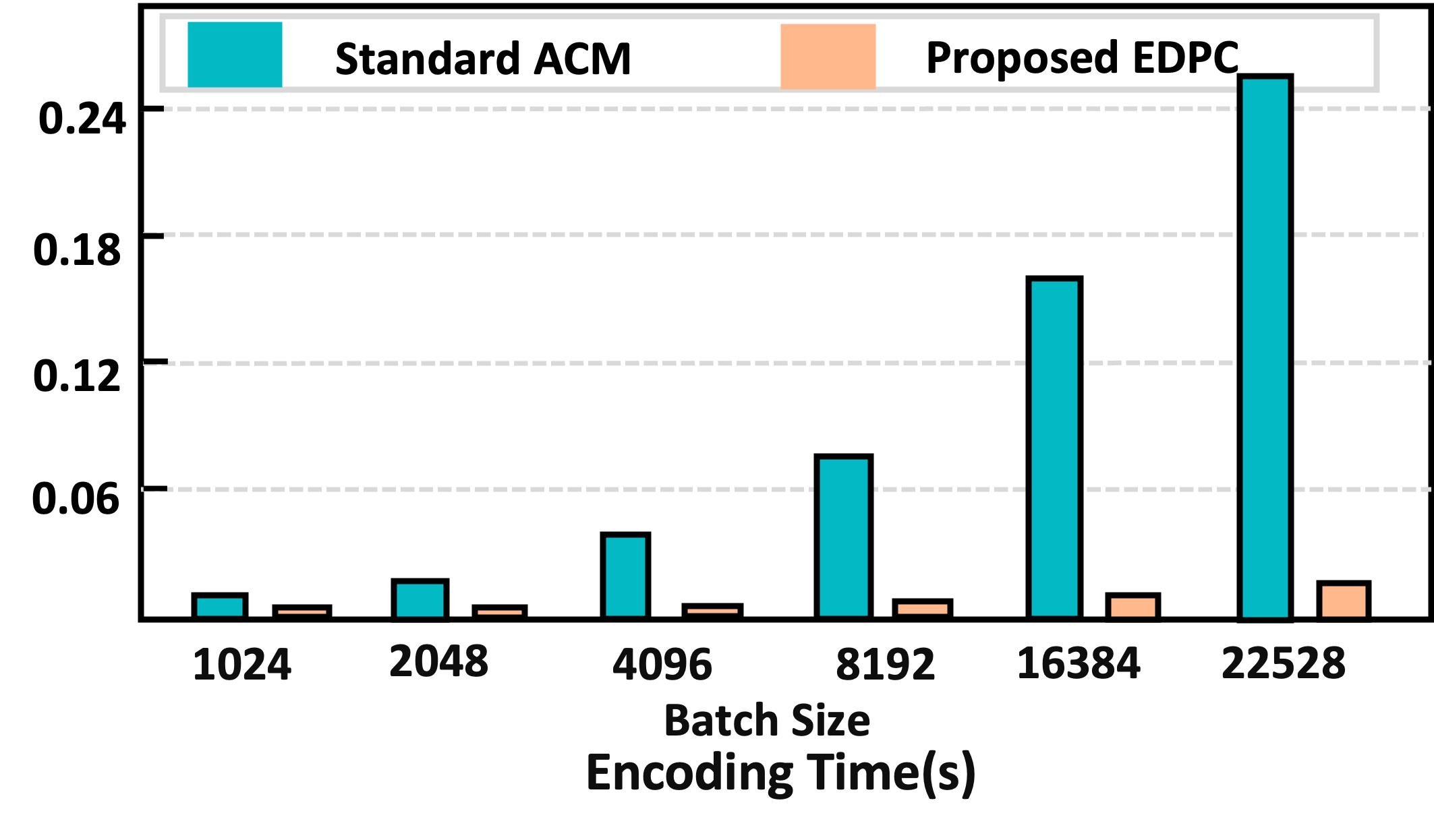}
    \setlength{\abovecaptionskip}{-0.4cm} 
\caption{Comparison of encoding time between \modelname{} and the standard ACM across different batch sizes.}
\label{enc}
\vspace{-0.5cm}
\end{figure}

We evaluate the impact of the Decoupled Pipeline Compression Architecture (DPCA) on encoding efficiency. As shown in the last row of Table~\ref{tab:ablation_compression}, removing the pipeline component results in a significant drop in speed, from 157~KB/s to 70~KB/s (batchsize = 4096)—indicating that DPCA boosts compression speed by 2.24$\times$. Furthermore, Figure~\ref{enc} illustrates the benefit of multi-process encoding within DPCA. Across varying batch sizes, \modelname{} achieves a substantial reduction in encoding time, with a maximum observed speedup of 21.73$\times$, approaching an order-of-magnitude improvement. These results highlight DPCA's critical role in enabling high-throughput, low-latency compression.

\section{Conclusion}

In this work, we present \modelname{}, an Efficient Dual-path Parallel Compression framework designed to address the limitations of existing autoregressive compression models. At the modeling level, \modelname{} introduces a multi-branch architecture guided by the proposed Information Flow Refinement (IFR) metric, effectively enhancing byte-level feature diversity through the Multi-path Byte Refinement Block (MBRB). At the system level, the Latent Transformation Engine (LTE) significantly reduces memory and parameter overhead, while the Decoupled Pipeline Compression Architecture (DPCA) enables pipelined parallelism with multi-process encoding to accelerate compression.
Extensive experiments demonstrate that \modelname{} achieves state-of-the-art compression ratios while substantially improving runtime efficiency and resource usage. Notably, it reduces model parameters by up to 4×, lowers GPU memory consumption by 1.91×, accelerates compression speed by 2.7×, and improves compression ratio by 3.2\% compared to strong baselines. These results confirm the effectiveness and practicality of \modelname{} in large-scale, real-time multimedia compression scenarios.

% \section{Acknowledgments}
% \begin{acks}
% acknowledgements
% \end{acks}

%\section{Supplementary Material}

\begin{acks}
This work is supported in part by the National Natural Science Foundation of China under grant 62171248, 62301189, and Shenzhen Science and Technology Program under Grant KJZD20240903103702004, JCYJ20220818101012025, GXWD20220811172936001.
\end{acks}

%\clearpage
\bibliographystyle{ACM-Reference-Format}
\bibliography{main}

%\appendix
%\input{sections/appendix}

\end{document}